\documentclass[twocolumn,floatfix,amsmath,prl,aps,showpacs]{revtex4}

\usepackage{color}
\usepackage{graphicx}
\usepackage{dcolumn}
\usepackage{bm}

\begin{document}

\title{A recursive method for the 
density of states in 
one dimension}

\author{Imke Schneider}
\affiliation{Department of Physics and Research Center OPTIMAS, University of Kaiserslautern, D-67663 Kaiserslautern, Germany}

\author{Sebastian Eggert}
\affiliation{Department of Physics and Research Center OPTIMAS, University of Kaiserslautern, D-67663 Kaiserslautern, Germany}

\date{\today}

\begin{abstract}
We derive a powerful yet simple method for analyzing the 
local density of states in gapless one dimensional fermionic systems, 
including extensions such as momentum dependent interaction parameters
and hard-wall boundaries.  
We study the crossover of
the local DOS from individual density waves 
to the well-known
asymptotic powerlaws and identify characteristic signs
of spin charge separation in possible STM experiments. 
For semi-infinite systems a closed analytic expression is found 
in terms of hypergeometric functions.
\end{abstract}                                                                 

%61.46.+w Nanoscale materials: clusters, nanoparticles, nanotubes, and nanocrystals
%68.37.Ef Scanning tunneling microscopy (including chemistry induced with STM)
%68.65.-k Low-dimensional, mesoscopic, and nanoscale systems: structure and nonelectronic properties
%71.10.Pm Fermions in reduced dimensions (anyons, composite fermions, Luttinger liquid, etc.)
%71.27.+a Strongly correlated electron systems; heavy fermions
%73.21.-b Electron states and collective excitations in multilayers, quantum wells, mesoscopic, and nanoscale systems
%73.21.Hb Quantum wires
%73.40.Gk Tunneling (for tunneling in quantum Hall effects, see 73.43.Jn)
%73.23.-b Electronic transport in mesoscopic systems
%73.63.-b Electronic transport in nanoscale materials and structures

\maketitle

The density of states (DOS) is a central quantity in the study of electronic condensed matter systems. 	The corresponding expression for the local DOS
\begin{eqnarray}
\rho(\omega,x)& = & 
\sum_{m} |\langle \omega_m| \Psi^\dagger(x)|0\rangle|^2 \delta(\omega-\omega_m)
\nonumber \\
&  = &   \frac{1}{2\pi}\int_{-\infty}^\infty e^{i\omega t}\langle\Psi^{}(x,t)
\Psi^{\dagger}(x,0)\rangle  dt
\label{ldos}
\end{eqnarray}
describes the probability of inserting an electron 
at a given energy $\omega$ and can be probed in 
tunneling experiments or, averaged over a range $x$, in photoemission experiments.
In two or more dimensions the DOS is typically peaked at renormalized
single particle excitation energies. 
In one dimensional 
systems, however,  interaction effects are enhanced, so that the DOS is
determined by collective many-body states instead \cite{Voit1995review}.
Accordingly, there has been considerable theoretical interest in analyzing 
the one dimensional DOS for several 
decades \cite{Voit1995review,Luther1974,Schoenhammer,Voit-Schoenh,
EggertMattsson,EggertAnfuso,Kivelson}.
Typical features that have been predicted 
are 
separate spin and charge density excitations and a characteristic
depletion of the DOS at low energies and near boundaries,
which have been 
seen experimentally in some special cases, such as cleaved edge overgrowth 
wires \cite{Auslaender}, 
superstructures on surfaces \cite{Segovia1999}, and
carbon nanotubes \cite{Bockrath1999}.
Recent theoretical activities in the field have produced notable 
advances in the areas of non-linear corrections \cite{Pustilnik2006},
numerical simulations \cite{Benthien2004, Schneider2008}, and applications of exact methods \cite{Pereira2009}.

On the other hand, there remain a number of open questions, 
especially in regards to the applicability of typical effective low energy theories 
such as the Luttinger liquid formalism to 
more realistic systems and models.
In fact a number of 
energy and length scales can affect the behavior, including
perturbations from other degrees of freedom, longer range electron-electron interactions,
impurities, and finite system sizes, so 
that realistic systems are never truly scale invariant.  
The renormalization due to higher order perturbing operators
leads to an energy 
and momentum dependence of the interaction parameters,
so that a description in terms of a single constant Luttinger liquid parameter 
is in general {\it not} adequate.
Accordingly, the low energy theory is altered substantially and 
the generically predicted powerlaw behavior  
with energy, momentum, and position is changed
into a more complicated behavior of the DOS.

In this paper, we address the
question of how to generally calculate the local DOS 
including complications which make the central interaction parameters effectively
energy and momentum dependent.  The calculation is based on the  
Fourier transform in Eq.~(\ref{ldos}) for vertex correlation functions
in  finite systems, which 
yields an expansion in delta-functions
 for the discrete DOS with 
coefficients that follow a recursion relation for
arbitrary momentum dependent interaction parameters.
It is shown that 
boundaries cause a natural scale dependence in the description of the
local DOS, which leads to the crossover from boundary to bulk behavior. 
Using a continuous description, we obtain a closed analytic expression
of the local DOS as a function of energy and position.
In order to identify spin- and charge separation by scanning tunneling microscopy (STM)
it is useful to analyze the spatial Fourier transform of the local DOS. 

The starting point is the general
expression of chiral fermionic fields in terms of
 vertex operators in the normal ordered form
\begin{eqnarray}\label{operator}
O^\dagger(x,t):=c \ 
e^{i \sum_{\ell=1}^{\infty}\frac{1}{\sqrt{\ell}} e^{i \ell \Delta \omega t}A_\ell^\dagger(x)}
e^{{i \sum_{\ell=1}^{\infty}\frac{1}{\sqrt{\ell}} e^{-i \ell \Delta \omega t}A_\ell(x)}},
\end{eqnarray}
where $A_\ell$ are 
linear combinations of bosonic annihilation operators, e.g.~of the form
$A_\ell(x)=\alpha_\ell e^{ik_\ell x} b_\ell^R + \beta_\ell e^{-ik_\ell x} 
b_\ell^L$ 
for a periodic system with length $L$, energy spacing $\Delta \omega =\frac{2\pi v}{L}$,
and $k_\ell = \frac{2 \pi \ell}{L}$ \cite{reveggert}.
Such operators $O^\dagger$ are used to represent left- or right moving 
fermion operators $\psi_{L/R}^\dagger$. 
In the case of several fermion channels (e.g.~spin and charge) the chiral
fermion field will be the product of operators $O^\dagger(x,t)$ 
for each channel separately, all
 of which have the form in 
Eq.~(\ref{operator}), but with different energy spacings $\Delta \omega$.
The magnitude of the 
prefactor $c$ 
is typically unknown unless 
a comparison with exact results can be made.
Zero mode terms have been omitted in Eq.~(\ref{operator})
since they only shift the spectrum of the DOS.

The form of the vertex operators in Eq.~(\ref{operator}) is 
believed to apply to gapless interacting 
fermion systems in one dimension. However, it is important to note that 
the expression (\ref{operator}) already implies 
that the possible energy levels are assumed to be 
evenly spaced at $\omega =\ell \Delta \omega$ relative
to the Fermi energy $\epsilon_F$,
which  is generally {\it not} exactly justified in realistic
systems and is also explicitly violated in any finite lattice model.
The reason why bosonization remains useful is 
that the central information 
about the electron-electron interactions is encoded in the Bogoliubov rotations, 
i.e.~the exact form of the linear combinations of 
bosons $A_\ell(x)$ in Eq.~(\ref{operator}). 
These do not crucially 
depend on the assumption of an equally spaced spectrum, since they are 
derived in momentum space.  Therefore, the typical {\it bosonization theory
does not describe the exact energy locations of the levels
 very accurately, but it appears to predict the
spectral weights  surprisingly well}, as can also be seen by 
numerical studies \cite{Schneider2008}.  

The 
general time correlation function can be calculated from Eq.~(\ref{operator}) 
\begin{eqnarray}\label{correlation_general}
\langle O(x,t) O^{\prime\dagger}(x,0) \rangle=|c|^2\exp 
(\sum_{\ell=1}^{\infty}\frac{1}{\ell} e^{-i \ell \Delta \omega t}\gamma_\ell(x)) ,
\end{eqnarray}
where $\gamma_\ell(x)=[A_\ell^{\phantom{\dagger}}(x),A_\ell^{\prime\dagger}(x)]$.
In accordance with the periodicity in time,
the Fourier transform in Eq.~(\ref{ldos}) yields
an expansion in delta-functions for the DOS
\begin{eqnarray}\label{integral_1}
\frac{1}{2\pi}\int_{-\infty}^{\infty}dt\, e^{i \omega t}\langle O(t) 
O^{\prime\dagger}(0) \rangle =\sum_{m} \rho_{m} \, \delta(\omega-m \Delta \omega).
\end{eqnarray}
By partial integration a 
recursion formula for the individual spectral weights $\rho_m$ is obtained
\begin{eqnarray}\label{recursion}
\rho_m= \frac{1}{m}  \left(\rho_{m-1} \gamma_1 + \rho_{m-2}\gamma_2 + ... 
+ \rho_0 \gamma_m \right)
\end{eqnarray}
%\sum_{\ell=0}^{m-1} \rho_\ell \gamma_{m-\ell}
with  $\rho_0=|c|^2$.
This simple but central result is the main instrument for analyzing the DOS 
in what follows. 
Besides the obvious simplicity of calculating the DOS with Eq.~(\ref{recursion}), it
has the major advantage of being free from divergences and regularization schemes.
Moreover, it is possible to analyze a dependence of the $\gamma_\ell$
on the principle energy quantum 
number $\ell$ which will be especially useful for boundaries and long range interactions.

Let us first re-examine the simplest example of a Luttinger liquid, 
which is a short-range interacting
single channel system with periodic boundary conditions. 
The commutator 
$\gamma=[A_\ell,A_\ell^\dagger]=\frac{1}{2}\left(\frac{1}{K}+K\right)$ 
is then given in terms of the 
Luttinger parameter $K$ independent of $\ell$.
In this case, Eq.~(\ref{recursion}) is solved analytically by
\begin{eqnarray}\label{solution_const}
\rho_m=|c|^2 \frac{\Gamma(m+\gamma)}{\Gamma(\gamma)\Gamma(m+1)} \approx |c|^2 \frac{1}{\Gamma(\gamma)}m^{\gamma-1}.
\end{eqnarray}
This is a well known result 
\cite{Schoenhammer}, which will be useful later.
However, 
many states can potentially contribute
to each spectral weight $\rho_m$, the 
number of which increases with the partitions of $m$.
As mentioned above, those states are in general 
not exactly degenerate at the energy $m\Delta \omega$. 
Therefore, the weights $\rho_m$ are spread in energy \cite{Bortz2009}, 
which is also the case in all examples that  follow.
However, the total value of $\rho_m$ is expected to be
rather accurate \cite{Schneider2008}.

Next, we consider fermions in a finite system 
with hard-wall   boundary conditions $\Psi(0)=\Psi(L)=0$. 
In this case, the operators in Eq.~(\ref{operator})
are of the form
$A_\ell(x)=(\alpha_\ell e^{ik_\ell x} + \beta_\ell e^{-ik_\ell x}) b_{\ell}^R$
with $k_\ell = \frac{ \pi \ell}{L}$ \cite{Eggert1992}.
The Green's function splits into a uniform and 
a $2 k_F x$ oscillating part
$\langle \Psi^{}(x,t) \Psi^{\dagger}(x,0)\rangle  = (G_R(x,x,t)
-e^{i 2k_F x} G_R(x,-x,t)) +c.c.$,
where $G_R(x,y,t) = \langle \psi_{R}^{\phantom\dagger}(x,t)\psi_{R}^{\dagger}(y,0) \rangle$ 
\cite{EggertMattsson,EggertAnfuso,Kivelson,Eggert1992}.
For spinful fermions  those correlation functions have the form of 
Eq.~(\ref{correlation_general}) for the 
spin factor ($K_s,\ v_s$) and the charge factor ($K_c,\ v_c$) separately 
\begin{eqnarray}
G_R(x,x,t)
=
|c_x|^2
\prod_{\nu=c,s} 
\exp({\sum_{\ell=1}^\infty \frac{1}{\ell}e^{-i\ell\Delta\omega_\nu
t}\gamma_{\nu,\ell}^{\rm{uni}}(x) })
\end{eqnarray}
\begin{equation}
G_R(x,-x,t)
=|c_x|^2 \prod_{\nu=c,s} 
\exp({\sum_{\ell=1}^\infty \frac{1}{\ell}e^{-i\ell \Delta\omega_\nu
t}\gamma_{\nu,\ell}^{\rm{osc}}(x)}),
\nonumber 
\label{cfopen}\end{equation}
where the commutators are dependent 
on position $x$ and mode $\ell$ in this case
$\gamma_{\nu,\ell}^{\rm{uni}}(x)= \frac{1}{4 K_\nu}+\frac{K_\nu}{4} +
(\frac{1}{4 K_\nu}-\frac{K_\nu}{4}) \cos(2 k_\ell x) $ and
$\gamma_{\nu,\ell}^{\rm{osc}}(x)= (\frac{1}{4 K_\nu}-\frac{K_\nu}{4})
+\left(\frac{1}{4K_\nu}+\frac{K_\nu}{4}\right)
\cos(2 k_\ell x) + \frac{i}{2}\sin(2 k_\ell x)$
with $\Delta \omega_\nu= v_\nu \frac{ \pi }{L}$ for $\nu=s,c$.
Also the prefactor is position dependent
$|c_x|^2\propto\prod_\nu (\sin \frac{\pi x}{L})^{\frac{1}{4K_\nu}-\frac{K_\nu}{4}}$ from 
normal ordering.  Using 
Eq.~(\ref{recursion}) for the
uniform/oscillating and spin/charge parts separately
$\rho^{\rm{uni/osc}}_{\nu,m}=\frac{1}{m}\sum_{\ell=1}^{m}\,\rho^{\rm{uni/osc}}_{\nu,m-\ell}
\gamma_{\nu,\ell}^{\rm{uni/osc}}$, 
it is possible to solve the 
Fourier transform (\ref{ldos}) even for this more complicated case.
The local DOS at the energy $\omega = m_c \Delta \omega_c +m_s \Delta \omega_s$
is then simply the folded product of spin and charge, i.e.
\begin{equation}\label{ldos_open}
\rho_{\rm{u/o}}=\!\sum_{m_c,m_s}\!\rho^{\rm{uni/osc}}_{c,m_c}\,\rho^{\rm{uni/osc}}_{s,m_s}
\delta\left(\omega\!-\!m_c \Delta \omega_c\! -\! m_s \Delta \omega_s \right)
\end{equation}
and 
$\rho(\omega,x)=(\rho_{\rm{u}}(\omega,x)-e^{i2k_Fx}\rho_{\rm{o}}(\omega,x)) + c.c.$.

\begin{figure}
   \begin{center}
        \includegraphics[width=.49\textwidth]{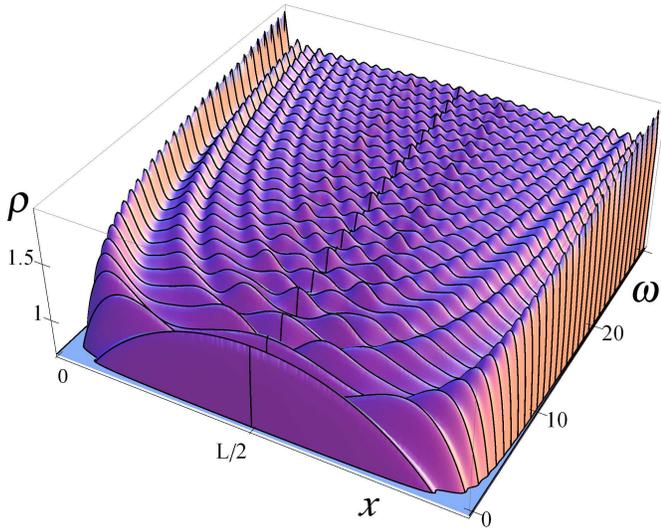} 
   \end{center}
    \caption{(Color online) $\rho_{\rm{u}}(\omega,x)$ in units of $|c_{x=L/2}|^2$ 
for $K_c=0.531$, $K_s=0.618$ and $\upsilon_c/\upsilon_s=1.618$ averaged over an 
energy interval of  $0.3$ in units of $\Delta \omega_c$. }
   \label{realspace}
\end{figure}

Equations (\ref{recursion}) and (\ref{ldos_open})
 can be evaluated very quickly with a few lines of 
code for a large number of levels and different parameters. 
The analytic expressions for each level are 
superpositions of density waves up to 
wavenumbers corresponding to $m_s$ and $m_c$. 
The level density increases with $\omega$, 
according to the number of ways of choosing 
$m_s$ and $m_c$ for a given energy interval in Eq.~(\ref{ldos_open}).
It is therefore useful to average over a small 
energy window, 
which results in a typical local DOS as a function of $x$ and $\omega$ as shown in 
Fig.~\ref{realspace} for $K_c=0.531$, $K_s=0.618$ and $v_c/v_s =1.618$, where the 
oscillating part has been omitted.
The solid lines in Fig.~\ref{realspace} show the characteristic standing density waves 
with higher wavenumbers as the energy increases,
with scaling of features
along hyperbolas with $\omega = m v_c/x$, while $v_s$ plays a lesser role.
In contrast, 
for non-interacting systems $K_c=K_s=1$ and Fig.~\ref{realspace} 
is completely flat.

\begin{figure}
   \begin{center}
        \includegraphics[width=.49\textwidth]{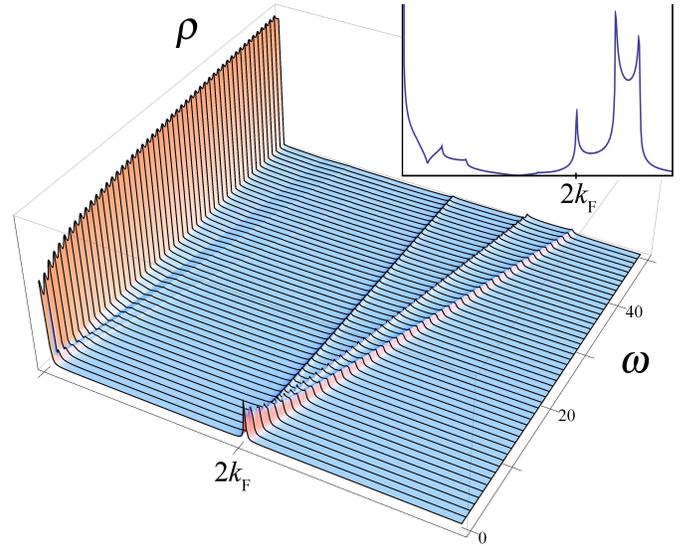}
   \end{center}
    \caption{(Color online) Spatial Fourier transform of the local DOS with parameters 
as in Fig.~\ref{realspace} in arbitrary units. Inset:  Cut at fixed energy.}
   \label{fourier1}
\end{figure}

In order to get clear evidence of the spin and charge separation it is 
necessary to consider the oscillating part, which is best analyzed by a 
Fourier transform as first considered in Ref.~[\onlinecite{Kivelson}].
In Fig.~\ref{fourier1} we have plotted the absolute value of the Fourier
transform in space of the local DOS including uniform and oscillating parts. 
It is possible to identify separately dispersing 
spin and charge features starting at $2k_F$.  In addition there is a third 
maximum at $2k_F$ for all energies.  The uniform DOS at $k=0$ shows the typical powerlaw
increase. 
For  spin-independent interactions $K_s=1, K_c<1$ the results for Figs.~\ref{realspace} 
and \ref{fourier1} look qualitatively 
very similar (not shown), except for a missing slowly dispersing maxima which can barely be detected
in Fig.~\ref{fourier1} starting from $k=0$.

\begin{figure} 
   \begin{center}
        \includegraphics[width=.49\textwidth]{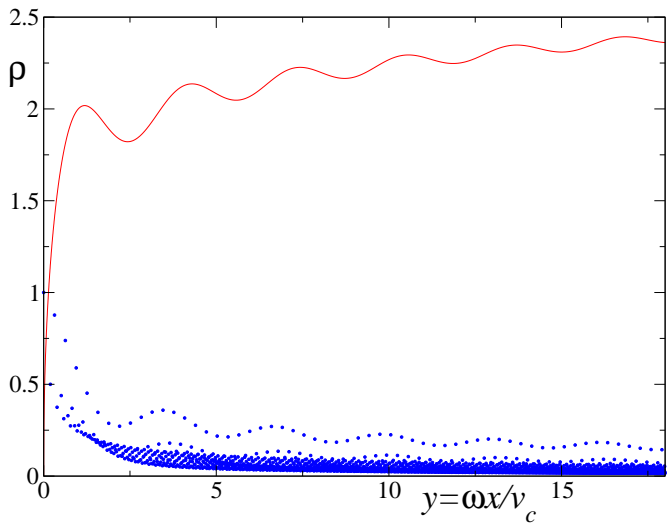}
   \end{center}
\caption{(Color online) The uniform spectral weights $\rho_{\rm u}$ in units of $|c_x|^2$
 near a boundary
as a function of $\omega x/v_c$ for $K_c=0.531$, $K_s=1$, 
and $\upsilon_c/\upsilon_s=1.618$.  The large number of spin and charge levels for $x=0.1L$
(blue dots)
follow Eq.~(\ref{hypergeo}) (red line) after averaging.}
\label{dos}
\end{figure}

A further analytical analysis of the local DOS 
is possible in the continuous limit of the level spacing
$x m_\nu \gg L $. 
In this case, Eq.~(\ref{recursion}) becomes an integral equation
\begin{eqnarray}\label{integraleqn}
\omega \,\rho(\omega,x)= \int_0^\omega  \rho(\omega^\prime ,x)\,\gamma (\omega-\omega^\prime,x) \;d \omega^\prime ,
\end{eqnarray}
which again holds for the oscillating/uniform and spin/charge parts
separately, where $\gamma (\omega,x)$ are the corresponding continuous
functions. 
Let us focus on the uniform charge part close to a single edge, i.e.
$\gamma(\omega, x)= \frac{1}{4 K_c}+\frac{K_c}{4} +\left(\frac{1}{4 K_c}-\frac{K_c}{4}\right) \cos(\frac{2 \omega x}{v_c}) $
from above.
After rewriting Eq.~(\ref{integraleqn}) in terms of the dimensionless
variable $y = \omega x/v_c$, differentiating three times, and using the 
fact that $\gamma'' = 4(\frac{1}{4 K_c}+\frac{K_c}{4}-\gamma)$, it is possible to obtain
a third order differential equation, which in turn can be solved by 
the following hyper\-geo\-metric function
\begin{equation}
\rho^{\rm{uni}}_c(y)=
y^{\frac{1}{2 K_c}-1}  \,_1\!F_2\left(\tfrac{1}{8 K_c}-\tfrac{K_c}{8};\tfrac{1}{4 K_c},\tfrac{1}{4 K_c}+\tfrac{1}{2},-y^2\right),
  \end{equation}   
with $ _1\!F_2 (a;b_1,b_2,z)=\frac{\Gamma(b_1)\Gamma(b_2)}{\Gamma(a)}\sum_{k=0}^\infty 
\frac{\Gamma(a+k)}{\Gamma(b_1+k)\Gamma(b_2+k)}\frac{z^k}{k!}$. The dependence on
position is not yet specified. 
The full local DOS from several channels 
can then be obtained by folding according to the continuous version of 
Eq.~(\ref{ldos_open}) which for $K_s=1$ again results in a hyper\-geo\-metric 
function 
\begin{eqnarray}
\rho_{\rm u}&= & f(x) \tfrac{v_c}{v_s}
\int_0^y \,\rho^{\rm{uni}}_c(y') \rho^{\rm uni}_s\left(\tfrac{v_c}{v_s}(y-y')\right)dy' \\
\label{hypergeo}
&=&\sqrt{\frac{v_c}{v_s}}\frac{|c_x|^2}{\Gamma{(\tfrac{1}{2}+\frac{1}{2 K_c})}}
\left(\frac{\omega}{\Delta\omega_c}\right)^{\tfrac{1}{2 K_c}-\tfrac{1}{2}} \nonumber \\
& & \times
 \,_1\!F_2\left(\tfrac{1}{8 K_c}-\tfrac{K_c}{8};
\tfrac{1}{4 K_c}+\tfrac{1}{4},\tfrac{1}{4 K_c}+\tfrac{3}{4},-(\tfrac{\omega x}{v_c})^2
\right)  \nonumber 
\end{eqnarray}
as shown in Fig.~\ref{dos}.
The $x$-dependence $f(x)$ is fixed by the asymptotic behavior near  $x=0$ which yields 
the expression (\ref{solution_const}) with $\gamma= 
\frac{1}{2K_c}+\frac{1}{2}$. 
Therefore an exact analytic formula (\ref{hypergeo}) has been derived
without the need to 
evaluate the Fourier transform of correlation functions
in Eq.~(\ref{ldos}), which involves a complicated contour
integration \cite{EggertMattsson} and has so far not 
been possible analytically.
Figure \ref{dos} also shows the DOS for all $m_s$ and $m_c$ quantum numbers
from the discrete recursion formula (\ref{recursion}).  Clearly, 
the individual weights {\it drop} with increasing $\omega$, but the averaged 
behavior follows the analytical prediction (\ref{hypergeo}).  
The boundary and bulk exponents of the effective powerlaws $\rho \propto \omega^\alpha$ are
given by $\alpha \sim \sum_\nu \gamma^{\rm uni/osc}_\nu-1$ expanded for small
$x$ or averaged over large $x$, respectively,  which are however only rough approximations
compared to the analytic expressions above.

So far, we have considered a constant Luttinger parameter $K$, where the scale dependence
of the DOS stems only from the boundary.  However, for longer range interactions,
$K_\ell$ itself is expected to be dependent on $\ell$ even in the bulk,  
e.g.~$K_\ell =  1/\sqrt{V_0 \ln \frac{a}\ell}$ for the case of a
Coulomb potential \cite{schulz}.
There has been a controversy in the literature about the behavior of the 
corresponding DOS, for which 
an effective exponent $\rho\propto \omega^{\alpha(\omega)}$
was postulated in Ref.~[\onlinecite{coulomb1}], which in turn lead to 
a debate \cite{coulomb1,coulomb2}.  Our analysis with 
$\gamma_\ell=\frac{1}{2}\left(\frac{1}{K_\ell}+K_\ell\right)$ 
using Eq.~(\ref{recursion}) now shows that
over a limited range 
the true behavior can fit to the 
form postulated in Ref.~[\onlinecite{coulomb1}] by some choice of parameters, 
but that there is an additional unknown scale dependence as the authors also 
mention \cite{coulomb1}.  Therefore, the description in terms
of $\alpha(\omega)$ is not complete, but phenomenologically useful.
An analysis with help of Eq.~(\ref{recursion}) shows that  corrections 
come from derivatives in $\gamma_\ell$, which however
cannot be summed exactly.  Using the recursion formula it is 
possible to 
consider interactions $K_\ell$ of arbitrary form with relative ease.
Also for harmonically trapped fermions a description of a mode dependent
parameter $K_\ell$ has been predicted, albeit with a yet unknown dependence
on $\ell$ \cite{harmonic}.

This brings us to an important application of the 
recursion formula for systems which are {\it not} perfect Luttinger liquids.
As outlined in the introduction, potentially all realistic models contain 
some scale-dependence of the parameter $K_\ell$,
which is however not a priori known. 
%Even for short-range interacting models
%like the Hubbard model \cite{logcorrections} it is well known that logarithmic 
%or higher order corrections lead to a scale dependent renormalization, 
Therefore, it will be very beneficial to use the recursion formulas (\ref{recursion})
in reverse:  Knowing the lowest $m$ spectral weights $\rho_\ell$, it is easily possible 
to uniquely determine the first $m-1$ interaction parameters $\gamma_\ell$.  
The analogous statement is also true for continuous spectra.  
This is especially promising considering the 
fast progress in 
numerical \cite{Benthien2004, Schneider2008}, 
exact \cite{Pereira2009}, 
and experimental \cite{Auslaender,Segovia1999,Bockrath1999} techniques in determining the DOS.
 
In summary, we have obtained a straight-forward tool for analyzing the local DOS
in one dimensional interacting systems.  The crossover from individual levels 
to a continuous spectrum has been studied in detail for future comparison with 
possible STM experiments.
A closed analytic expression was obtained for electrons in a system with
hard-wall boundaries.  Long range interactions can also be considered.  
The inversion of the recursion formula is useful to calculate 
the interaction parameters directly from the low energy spectral weights, 
which in turn can be obtained from numerical, exact, or experimental methods.
In this way the theory can be generalized to systems which are not perfect Luttinger
liquids, such as harmonically trapped ultra-cold fermions.

\begin{acknowledgments}
 We are thankful for useful discussions with I.~Affleck,
M.~Bortz, H.~Johannesson, V.~Meden, D.~Schuricht, S.~S\"offing, and A.~Struck.
This work was supported by
the DFG and the State of Rheinland-Pfalz via
the SFB/Transregio 49 and the MATCOR school of
excellence.

\end{acknowledgments}

\end{document}